# Reconfigurable Distributed FPGA Cluster Design for Deep Learning Accelerators

Hans Johnson, Tianyang Fang, Alejandro Perez-Vicente and Jafar Saniie
*Embedded Computing and Signal Processing (ECASP) Research Laboratory (http://ecasp.ece.iit.edu)*
*Department of Electrical and Computer Engineering*
*Illinois Institute of Technology, Chicago IL, U.S.A.*

*Abstract* — We propose a distributed system based on low-power embedded FPGAs designed for edge computing applications focused on exploring distributing scheduling optimizations for Deep Learning (DL) workloads to obtain the best performance regarding latency and power efficiency. Our cluster was modular throughout the experiment, and we have implementations that consist of up to 12 Zynq-7020 chip-based boards as well as 5 UltraScale+ MPSoC FPGA boards connected through an ethernet switch, and the cluster will evaluate configurable Deep Learning Accelerator (DLA) Versatile Tensor Accelerator (VTA). This adaptable distributed architecture is distinguished by its capacity to evaluate and manage neural network workloads in numerous configurations which enables users to conduct multiple experiments tailored to their specific application needs. The proposed system can simultaneously execute diverse Neural Network (NN) models, arrange the computation graph in a pipeline structure, and manually allocate greater resources to the most computationally intensive layers of the NN graph.

## I. Introduction

The field of Deep Learning (DL) has witnessed significant advancements due to collaborative research efforts in both Hardware (HW) and Software (SW) designs. Although DL frameworks have facilitated the exploration of novel DL architectures, Electronic Design Automation (EDA) tools have lagged behind and Register Transfer Level (RTL) designs still rely on traditional HDL. Despite improvements in C++/High-Level Synthesis (HLS) tools, the resulting RTL often consumes excessive logic resources and can be challenging to modify, leading to a growing gap between HW and DL architecture.

A primary challenge lies in supporting new operations on HW as Neural Network (NN) computation graphs become more complex. With ASIC Neural Processing Units (NPUs), Processing Elements (PE) are fixed, and DL compilers must support new computations on existing hardware—a complex and time-consuming process. As the demand for efficient DL computation at the edge increases, the focus has shifted towards optimizing NN architectures and allocating dedicated hardware to appropriate computational blocks for improved power efficiency, reduced latency, and scheduling optimizations. While DL frameworks have contributed to the progress of new architectures, the development of dedicated ASIC (Application Specific Integrated Circuit) hardware for DL workloads is more time-consuming due to EDA tool limitations and RTL development costs vs FPGA [1-3].

As workload intensity and compute resource requirements vary between applications, a more dynamic approach is necessary. FPGAs play a crucial role in providing adaptability and parallelism while maintaining low latency and optimal power levels. Our proposed FPGA cluster architecture interconnected through an Ethernet switch, functions as a hardware stack that accommodates diverse NN models, arranges computation graphs in a pipeline structure and allocates resources to computationally intensive layers of the NN graph. This versatile system effectively addresses the challenges of supporting new operations and handling varying workloads in edge computing applications.

The importance of exploring FPGA cluster architecture lies in its inherent advantages for edge computing applications in the realm of Deep Learning. FPGA clusters offer superior performance compared to traditional CPU/GPU-based systems regarding low latency applications. Low latency directly translates to faster data processing speeds that are essential for edge computing applications, especially in scenarios where instant decision-making is crucial such as autonomous vehicles, drones, and network configurations [4, 5]. These real-time applications require rapid processing and decision-making capabilities to ensure safety and efficiency. By utilizing FPGA clusters in edge computing applications, it becomes possible to achieve the required processing speeds and cater to the demands of latency-sensitive applications, providing a compelling reason for the exploration and development of FPGA cluster architectures.

As deep learning accelerators become increasingly essential for driving advancements in technology and edge computing, their compatibility with FPGA platforms offers a promising solution for harnessing the full potential of versatile and adaptable hardware-software co-designs. The Versatile Tensor Accelerator (VTA) is an open-source, scalable, and customizable deep learning accelerator designed to address the challenges of deploying deep learning workloads on a wide range of hardware platforms [6, 7]. VTA enables researchers and engineers to explore the co-design of both software and hardware to create efficient and adaptable deep learning systems. It provides a flexible and extensible infrastructure that supports a variety of neural network models and accelerates their computation.

VTA is designed to work seamlessly with popular deep learning frameworks like Apache TVM, an end-to-end compiler stack for deep learning systems. By integrating with these frameworks, VTA enables the efficient execution of deep learning workloads on various hardware platforms, including FPGAs, ASICs, and other custom accelerators. The modularity of VTA allows users to easily adapt and extend the accelerator design to accommodate the evolving requirements of deep learning algorithms and achieve the desired performance, power efficiency, and latency.

In section II we discuss the proposed system as an FPGA cluster in terms of the hardware, firmware, and software design. In section III we show our results from one development using VTA and discuss our results, current, and future research in section IV. Finally, we leave with concluding remarks in section V.

## II. SYSTEM DESIGN

### A. Hardware

The cluster hardware features two distinct FPGA SoC variants, both founded on the same heterogeneous system concept where a low-power Processing System (PS) connects to Programmable Logic (PL). The primary distinction between these FPGA SoCs lies in the available PL logic resources and the PS CPU performance. The compute-lite cluster incorporates up to 12 Xilinx Zynq-7020 chips, combining PYNQ-Z1 as well as ZedBoards. As only the Ethernet port and the Zynq-7020 chip are utilized, there's no need to consider the differences in I/O peripherals between the two boards. The Zynq-7020 is an All-Programmable System on Chip (APSoC) that seamlessly integrates an FPGA with a multi-core processor into a single, unified circuit. The programmable logic (PL) inside the APSoC consists of 13,300 logic slices (6 LUTs and 8 flip-flops), 630 KB fast block RAM, 220 DSP slices, and a 50 Hz input clock. In addition to the PL, the APSoC includes PS with a 650 MHz dual-core Cortex-A9 ARM processor, a DDR3 memory controller with 8 DMA channels, and 4 high-performance AXI3 slave ports for communication between the PL and PS. A picture of this FPGA stack can be observed in Fig . 1.

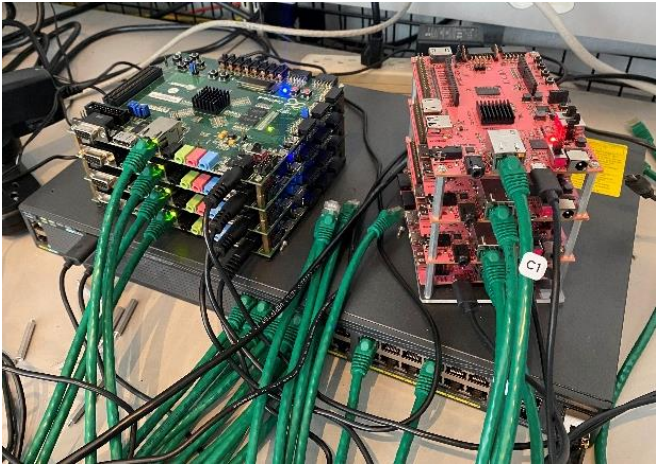

Fig. 1. FPGA Stack with Four ZedBoards and Four PYNQ-Z1s

For a more computationally intensive FPGA cluster, the proposed system will integrate 5 Zynq UltraScale+ MPSoC platforms by Xilinx into the stack, differing mainly on the number of logic units inside the chip from the Zynq-7000 boards. The MPSoC is based on a hybrid CPU-FGA architecture where the configurable logic and a multi-core processor are routed into a single chip. The programmable logic (PL) combines logic cells (LUTs and FFs), BRAM, URAM, and DSP slices. In addition to the PL, the MPSoC includes Processing System (PS) with a 1.5 GHz Quad-core Arm Cortex-A53 processor, a 600 MHz Dual-core Cortex-R5 RT processor, MaliTM-400 MP2 GPU, a memory controller with DMA channels, and high-performance AXI4 slave ports for communication between the PL and PS.

Zynq-7000 will provide fewer available resources than Zynq Ultrascale+ devices, also achieving lower clock rates if timing wants to be met without any negative slack or hold time violation. Additionally, the existing Arm CPU to the FPGA's fabric is comparably different, both in the instruction set and compute capacity. One of the main advantages of using Zynq-7000 is their overall power efficiency and cost, making it easy to scale computing for power constraint systems.

To connect all boards to the cluster, we used a standard Cisco switch together with RJ-45 connectors to connect the FPGA slave nodes to the master of the cluster. This switch offers 1GB/s ethernet speed. The system will be orchestrated from a master host PC although it could be done from one of the FPGA CPU nodes acting as master instead of the slave node.

### B. Firmware

In our design, VTA was explored because it offers a flexible and efficient solution to deep learning acceleration. The modular architecture allows for customizable and optimizable hardware parameters to suit specific application requirements. The VTA architecture is designed to optimize resource utilization and maximize performance in deep learning workloads. The VTA comprises four primary modules: fetch, load, compute, and store, which work together to enable high memory bandwidth usage for memory-bound workloads and efficient compute resource utilization on the PL side. A simplified block diagram for the VTA architecture can be observed in Fig. 2.

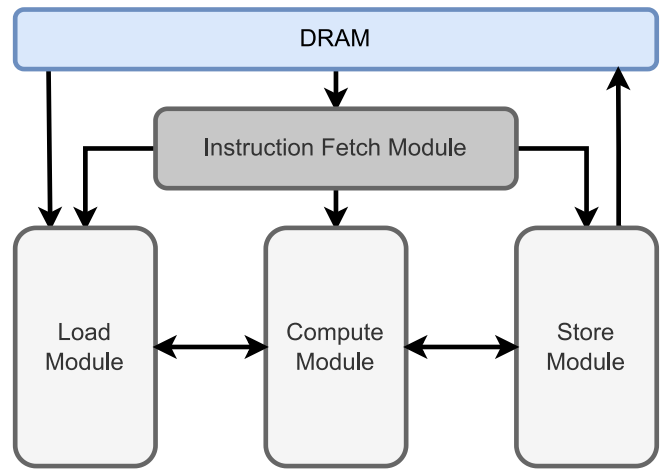

Fig. 2. Versatile Tensor Accelerator Block Diagram

On-chip memory SRAM is employed via unidirectional data channels to facilitate communication between the modules. Each of the four modules establishes a connection with both a consumer and a producer. By incorporating Read-After-Write (RAW) and Write-After-Read (WAR) queue dependencies, proper timing and execution of producer-to-consumer tokens are ensured. VTA architecture allows for the concurrent use of compute and memory modules to optimize resource usage in every clock cycle.

TVM achieves this optimization by generating virtual threads, which results in partitioning tasks into two separate execution contexts that prevent interference between fetch, load, compute, and store operations. VTA provides the capability to modify hardware parameters in the accelerator, including the GEMM (General Matrix Multiplications) core tensor intrinsic, I/O, weight, and accumulator tensor dimensions. Additionally, the on-chip SRAM port memory and data type sizes for weights and accumulation can be adjusted. VTA's register file can execute two tensor operators: The ALU handles element-wise tensor operations like addition, activation, pooling, and more, while GEMM performs more intricate arithmetic operations, such as complex matrix multiplication computations needed for 2D convolutions and dense layers.

The configuration parameters for the VTA involve mainly the on-chip memory buffer size and tensor data types, including the matrix multiply quantization and data type of input and parameter to perform computation inside the GEMM block. The initial hardware parameters followed the values in Table I below. Note that the clock speeds differ between the Zynq-7000 FPGA stack and the UltraScale+ FPGA stack. This is because meeting the timing requirements became the limitation of the VTA configuration.

Table I. Initial VTA Configuration Parameters

| PARAMETERS | SIZE |
|---|---|
| CLOCK_FREQUENCY (ZYNQ-7000) | 100 MHz |
| CLOCK_FREQUENCY (UltraScale+) | 300 MHz |
| INPUT_WIDTH | 8-bit |
| WIEGHT_WIDTH | 8-bit |
| ACCUMULATOR_WIDTH | 32-bit |
| BATCH_SIZE | 1 |
| BLOCK_SIZE | 16 |
| MICRO_OP_BUFFER_SIZE | 32 Kb |
| INPUT_BUFFER_SIZE | 32 Kb |
| WEIGHT_BUFFER_SIZE | 256 Kb |
| ACCUMULATOR_BUFFER_SIZE | 128 Kb |

*C. Software*

For the implementation of our programs, a PC acted as a control for the processes running on the FPGAs. This involved sending and retrieving data from the FPGAs which acted as accelerator cores in our experiments. FPGA-to-FPGA communication was not completely implemented for this configuration, but it could be easily done using AXI or AXIS protocols through Ethernet. We also used existing drivers for the software components for communication between the network interfaces and the PC.

When scheduling the NN computational graphs across the FPGA hardware domain, we tested four different approaches:

1. Scatter-Gather
2. AI Core Assignment
3. Pipeline Scheduling
4. Fused Schedule

Scatter-Gather is a technique used to increase the number of input images processed in a single inference run by distributing input frames across multiple FPGA channels. This method begins with a scatter operation to distribute data and ends with a gather operation to collect and store the outputs in an ordered batch. Scatter-Gather operations can occur at both ends of a DFG or in-between and can happen multiple times across the NN graph.

AI Core Assignment aims to maximize overall performance by assigning more compute resources to the bottleneck workload in the computational graph. This approach increases the number of consumer nodes for a given task and minimizes graph latency. It is crucial to maintain the order of subsequent computations on each assigned hardware to ensure tensors are gathered and processed correctly.

Pipeline Scheduling involves executing segments of an NN model in a distributed manner across independent or shared hardware resources. This method removes the single-input bottleneck by allowing the next input to be fed to each segment as soon as the consumer is free. As a result, all segments of the NN graph are consistently processing input data, increasing overall hardware utilization.

The Fused Schedule combines pipeline scheduling with AI core assignment to increase hardware utilization and distribute intensive compute tasks across the NN subgraphs. By allocating more compute units to the highest demanding segment, this approach reduces the NN bottleneck and continually performs computations across the subgraphs, maximizing the benefits of pipeline execution.

### III. RESULTS

The initial cluster test on the VTA platform was conducted using the parameters and clock frequency specifications outlined in Table 1. For this VTA configuration, a bitstream design without any timing violations or node overlaps was successfully generated for both the Zynq-7000 and UltraScale+ platforms.

Fig. 3 and Fig. 4 present the inference time required to process a single image through the ResNet-18 graph on the compute-lite cluster type on the Zynq-7000 FPGA stack and the UltraScale+ stack respectively. Fig. 3 (a) shows the table of execution time in milliseconds, and Fig. 3 (b) is a graphical representation of this data for the Zynq-7000 stack. The same is true for Fig. 4 (a) and (b), but for the UltraScale+ stack. The model was trained with an input shape of (N, 224, 224, 3), and no input resizing was applied during the inference step. The obtained values were categorized based on the number of compute resources utilized and the cluster strategy employed for distributing the NN workloads. Inference time results were determined by performing 10 evaluations on 10,000 random test images extracted from the ImageNet test dataset. For each evaluation, the average inference time was calculated and then averaged across the 10 evaluation results. Each run evaluation was carefully verified to ensure no data discrepancies or deviations from the expected time interval occurred.

Upon running inference on a single FPGA, an optimized micro-kernel generated through AutoTVM schedule exploration resulted in an inference time of 27.34 ms. As the number of FPGA resources increases for all cluster strategies, the workload became more distributed, and the expected inference time for each input image should have decreased. However, the table reveals that reduced latency is not always directly proportional to the addition of more FPGA devices to the cluster. Among the four strategies, distributing bottleneck operators (those necessitating greater computing power) across more FPGAs proved more effective as the number of FPGAs in the cluster increased. Notably, this approach negatively impacts latency when two or three FPGA nodes are used. The primary factor contributing to this performance loss is network bandwidth and processor involvement in transmitting data packet streams between two or more FPGA devices. The distributed cluster was tested using RJ-45 connectors with speeds of up to 1 GB/s, compounded by the FPGA CPU's need to DMA data buffers from the FPGA's logic and transmit them through the network to the next node, resulting in significant CPU handling overhead. Moreover, buffers are sent as blocking call MPI messages, which also affect the overall node message-passing handshake.

When testing the UltraScale+ FPGA cluster, the results for the first VTA configuration showed an improvement of approximately 6% compared to the Zynq-7000 cluster.

*Zynq-7000: Different Scheduling Methods Execution Time (milliseconds)*

| Num. FPGA / Cluster | Scatter-Gather Method | AI Core Assignment | Pipeline Scheduling | Fused Schedule |
|---|---|---|---|---|
| 1 | 27.34 | 27.34 | 27.34 | 27.34 |
| 2 | 17.53 | 36.85 | 20.43 | 19.32 |
| 3 | 12.33 | 28.32 | 15.59 | 16.87 |
| 4 | 7.87 | 20.31 | 11.29 | 9.13 |
| 5 | 6.44 | 15.40 | 9.03 | 7.37 |
| 6 | 5.66 | 9.63 | 7.33 | 6.62 |
| 7 | 4.78 | 4.55 | 5.93 | 4.92 |
| 8 | 3.94 | 3.98 | 4.22 | 4.01 |
| 9 | 3.17 | 2.46 | 3.88 | 3.45 |
| 10 | 2.84 | 2.11 | 3.22 | 2.94 |
| 11 | 2.71 | 1.93 | 2.94 | 2.74 |
| 12 | 2.58 | 1.84 | 2.62 | 2.66 |

(a)

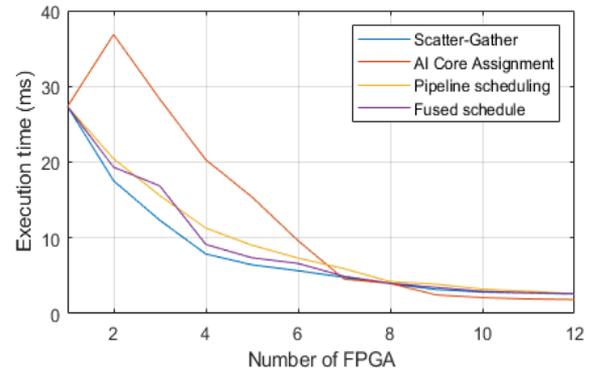

(b)

Fig. 3. Zynq-7000 VTA Cluster Test: a) Table of execution times in milliseconds per scheduling method, b) Execution time (ms) vs Number of FPGA

*UltraScale+: Different Scheduling Methods Execution Time (milliseconds)*

| Num. FPGA / Cluster | Scatter-Gather | AI Core Assignment | Pipeline scheduling | Fused schedule |
|---|---|---|---|---|
| 1 | 25.15 | 25.15 | 25.15 | 25.15 |
| 2 | 16.73 | 33.96 | 19.03 | 18.28 |
| 3 | 11.78 | 26.24 | 14.57 | 16.04 |
| 4 | 7.42 | 18.70 | 10.88 | 8.63 |
| 5 | 6.01 | 14.14 | 8.58 | 6.93 |

(a)

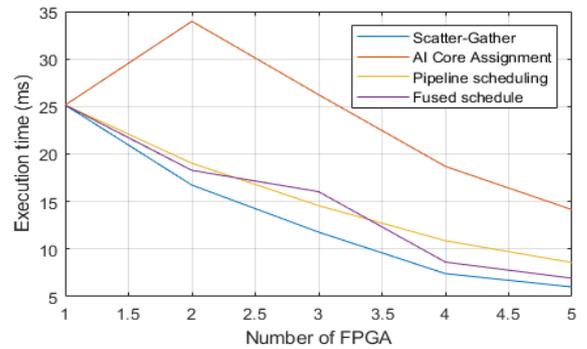

(b)

Fig. 4. UltraScale+ VTA Cluster Test: a) Table of execution times in milliseconds per scheduling method, b) Execution time (ms) vs Number of FPGA

## IV. DISCUSSION AND CURRENT RESEARCH

Our data provides evidence that a significant speedup with an FPGA stack for our tested DNN platform is exhibited. This correlates with current research that we experience significant improvements in throughput [8-11]. We also explored increasing the overall architecture parameters for the VTA platform in the UltlraScale+ stack without having timing violations in the RTL design. For the same configuration parameters as Table 1, we found the clock limit to be 350 MHz exhibiting a speedup of approximately 5.7% in execution time from Fig. 4. For another case we increased the GEMM block size to 32-bits, micro-op cache buffer, and input size to 64 Kb, weight buffer to 512 Kb, and accumulator to 256 Kb. The data type remained the same, and the clock frequency was reduced to 200 MHz to avoid negative hold slack. The result was a speedup of approximately 43.86% from Fig. 4.

For our experiment, our scope is limited by our one exploration of a DL Accelerator (DLA). In the future, we plan to implement several other DLA architectures on the FPGA cluster. Our targeted DLA architectures are Nvidia DLA, Tensil CU, PipeCNN, and Xilinx DPU.

## V. CONCLUSION

In conclusion, the implementation and evaluation of the VTA platform on Zynq-7000 and UltraScale+ FPGA clusters demonstrate the potential of leveraging FPGAs to accelerate deep learning workloads. The exploration of different cluster strategies, such as scatter-gather, AI core assignment, pipeline scheduling, and fused scheduling, has provided valuable insights into the trade-offs between latency, resource utilization, and network communication overhead. While the addition of FPGA resources in the cluster does not always result in a linear reduction of latency, distributing bottleneck operators across more FPGAs proves to be more effective as the number of FPGAs increases. Additionally, communication overhead and network bandwidth play crucial roles in determining the overall performance of the system. The findings from this study can serve as a foundation for future research in optimizing FPGA-based deep learning accelerators and their effective deployment in large-scale distributed systems.